\documentstyle[12pt, A4,epsfig]{article}
\thicklines                     
\begin{document}
\begin{titlepage}
\begin{centering}
\vspace{2cm}
{\Large\bf Time evolution of rolling tachyons}\\
\vspace{0.5cm}
{\Large\bf for a brane-antibrane pair}\\
\vspace{2cm}
Haewon~Lee\footnote{E-mail address: {\tt hwlee@hep.chungbuk.ac.kr}} and W.~S.~l'Yi\footnote{E-mail
address: {\tt wslyi@hep.chungbuk.ac.kr}} \vspace{0.5cm}

{\em Department of Physics}\\
{\em Chungbuk National University}\\
{\em Cheongju, Chungbuk 361-763, Korea}\\
\vspace{1cm}
\begin{abstract}
\noindent A precise form of the time evolution of rolling tachyons corresponding to a brane-antibrane
pair is investigated by solving the Hamiltonian equations of motion under the assumption that in a
region far away from branes the tachyon vacuum is almost already achieved, even at the beginning of the
rolling.
\end{abstract}
\end{centering}
\vspace{.5cm}
\vfill

Oct.~23, 2002\\
\end{titlepage}

\setcounter{footnote}{0}
%
\section{Introduction}
%
After the discovery of tachyon condensation\cite{Sen0} of non-BPS D-branes or brane-antibrane pairs,
much studies have been done to investigate the physics of unstable branes\cite{unstable-branes}. Some
possible roles of rolling tachyons on cosmology in relation to inflation are also discussed by many
authors\cite{cosmology}. Recently Sen proposed a new way to quantum gravity by interpreting tachyon
field as world lines of non-interating, non-rotating dust\cite{Sen1}. The important ingredient of his
proposal is that after sufficient rolling down of tachyon to vacuum, tachyon field is almost the same
as the time, thus making it possible to identify the elusive intrinsic time in general gravity with
tachyon field.

In this paper we use the Hamiltonian formalism\cite{{Sen1},{hamiltonian-formalism}} while solving the equations
of motion to obtain a precise form of time evolution of tachyon field. We first assume homogeneous
tachyon configuration in space, and then generalize it to a more realistic brane-antibrane
configuration.  In this case we take the ansatz that near the vacuum tachyon field $T$ can be written
as a summation of a function of time and a linear function of one of the spatial coordinate $x.$  It
will be proved in section 3 that this kind of ansatz is valid only if the tachyon potential around the
vacuum looks like $V(T) = \exp(-{1 \over 2}\alpha T)$ with some constant $\alpha.$ This is consistent
with string theoretic calculations that the tachyon potential should indeed be of that
form\cite{potential-form}. The potential around the region where D-branes reside is more complicated,
and this fact is related to the fact that it is not easy to guess any specific form of the potential
around $T\approx 0.$ But as long as one is far from the branes, one is able to have a precise form of
the time evolution of $T.$

In section 2, we briefly introduce the Hamiltonian formalism for the usual tachyon Lagrangian. The
canonical equations of motion are solved for homogeneous field configuration.  This way of solving the
equations of motion is more intuitive and readily applicable to more realistic systems such as
brane-antibrane pairs than the Lagrangian formalism\cite{Lagrangian-formalism}.  In section 3, we
assume a specific ansatz for tachyon field corresponding to a pair of brane-antibrane. By solving the
Hamiltonian equation of motion we obtain a precise form of time evolution of tachyon around the vacuum.
%
\section{Tachyon action and the equation of motion}
%
The Lagrangian density for tachyon field $T(x)$ on a space-filling D-brane system in a flat $p+1$
dimensional space is given by
\begin{equation}
{\cal L} = - V(T)\sqrt{1+\eta^{\mu\nu} \partial_\mu T
\partial_\nu T}
 = - V(T)\sqrt{ 1 + (\nabla T)^2 - (\partial_0 T)^2 },
 \label{cal-L}
\end{equation}
where $\eta^{\mu\nu} = {\rm diag}(-1, 1, \ldots ,1),$ and we use the convention $c=\hbar=\alpha'=1.$
The tachyon potential $V(T)$ is known to be maximum at $T=0,$ and vanishes as $T$ goes to infinity.
Instead of using the usual Lagrangian formalism for solving the equations of motion, we choose the
Hamiltonian formalism\cite{hamiltonian-formalism} which we briefly discuss.

First of all, we compare (\ref{cal-L}) with the usual Lagrangian of a relativistic particle of mass $m,$
which is given by
\begin{equation}
\int dt L = - m \int ds = - m \int dt \sqrt{ 1 - v^2}, \label{l}
\end{equation}
where $ds = \sqrt{-\eta_{\mu\nu} dx^\mu dx^\nu},$ and $v$ is the usual velocity.
When one rewrites (\ref{cal-L}) in the following form
\begin{equation}
{ {\cal L} \over \sqrt{ 1 + (\nabla T)^2}}  =
- V(T)\sqrt{ 1  - {(\partial_0 T)^2 \over  1 + (\nabla T)^2} },
 \label{L_of_T}
\end{equation}
it is clear that there are formal correspondences between tachyonic quantities and relativistic particle
quantities such as
\begin{eqnarray}
L &\rightleftharpoons& { {\cal L} \over \sqrt{ 1 + (\nabla T)^2}}, \label{lL} \\
m &\rightleftharpoons& V(T), \label{mV} \\
v &\rightleftharpoons& {\partial_0 T \over  \sqrt{1 + (\nabla T)^2}}. \label{vT}
\end{eqnarray}

Using the facts $p={\partial L \over \partial v},$ $\Pi(x) = {\partial {\cal L} \over \partial (\partial_0 T(x))},$
and the formal correspondences (\ref{lL}-\ref{vT}) one may infer that
\begin{equation}
p \rightleftharpoons \Pi.
\end{equation}
This means that from $p=mv / \sqrt{ 1 - v^2}$ one obtains
\begin{equation}
\Pi = {V(T)
\partial_0 T \over \sqrt{1- (\partial_0 T)^2 + (\nabla T)^2}}.\label{definition-of-Pi}
\end{equation}
In fact one may prove this from the very definition $\Pi(x) = {\partial {\cal L} \over \partial (\partial_0 T(x))}.$
Note that whenever $V(T) > 0,$ $\Pi(x)$ and $\partial_0 T$ have the same signs.

It is clear that the correspondence rule between $E=pv-L$ and the Hamiltonian density
${\cal H} = \Pi \partial_0 T - {\cal L}$ is the following,
\begin{equation}
E \rightleftharpoons { {\cal H} \over \sqrt{ 1 + (\nabla T)^2}}. \label{EH}
\end{equation}
This implies that the tachyonic counter part of the relativistic energy relation $E=\sqrt{p^2+m^2}$ is
\begin{equation}
{\cal H} = \sqrt{\Pi ^2 + V^2}\sqrt{1 + (\nabla T)^2}.\label{H}
\end{equation}
One may prove this directly from the definition of ${\cal H}$ and
\begin{equation}
\partial_0 T = {\Pi \over \sqrt{\Pi^2 + V^2} } \sqrt{1 + (\nabla
T)^2}\label{eq_T}.
\end{equation}
Here, (\ref{eq_T}) can be proven by using either (\ref{definition-of-Pi}) or $v = p / \sqrt{p^2 +m^2}.$
Another useful relation is the following,
\begin{equation}
\sqrt{1- (\partial_0 T)^2 + (\nabla T)^2} = {V \over \sqrt{\Pi^2 +
V^2} } \sqrt{1 + (\nabla T)^2},
\end{equation}
which is the tachyonic version of $\sqrt{1 - v^2} = m / \sqrt{p^2 +m^2}.$

Now we solve the equations of motion. Instead of solving the
usual hamiltonian equations of motion we solve $\partial_\mu T^{\mu}{}_{\nu} =0,$ where
the energy-momentum tensor is given by
\begin{equation}
T^{\mu}{}_\nu = -{\partial {\cal L} \over \partial \partial_\mu T} \partial_\nu T + \delta^{\mu}_{\nu} {\cal L}.
\end{equation}
By using the relations $T_{00} = {\cal H}$ and $T_{i0} = T_{0i} = \Pi \partial_i T,$ where $i \neq 0,$
it is easy to show that $\partial_\mu T^{\mu}{}_{0} =0$ reduces to
\begin{equation}
\partial_0{\cal H} = \partial_i(\Pi \partial_i T).\label{eq-H}
\end{equation}

To simplify the problem, we consider first the spatially homogeneous case such as $\partial_i T = 0$ for
all $i.$ In this case the funny factor $\sqrt {1 + (\nabla T)^2}$ which appears in (\ref{lL}), (\ref{vT}), and
(\ref{EH}) is just unity.  The equation (\ref{eq-H}) now reduces to
\begin{equation}
{d \over dt}{\cal H} = 0.
\end{equation}
This means that $\Pi^2 + V^2 = {\cal H}^2,$ which is the tachyonic form of $p^2 + m^2 = E^2,$ is
constant during the rolling of the tachyon. Using
(\ref{definition-of-Pi}), ${\cal H}^2$ can be written as
\begin{equation}
V^2{\dot{T}^2 \over 1 - \dot{T}^2} + V^2 = {\cal H}^2. \label{EM-relation}
\end{equation}
This is reminiscent of the energy-momentum relation,
\begin{equation}
m^2{v^2 \over 1 - v^2} + m^2 = E^2,
\end{equation}
of a relativistic particle of mass $m$ and velocity $v.$ The crucial difference is that for usual
particles masses are constant, but for tachyons $V(T)$ vanishes rapidly as time goes on.
\begin{figure}
\begin{center}
\psfig{figure=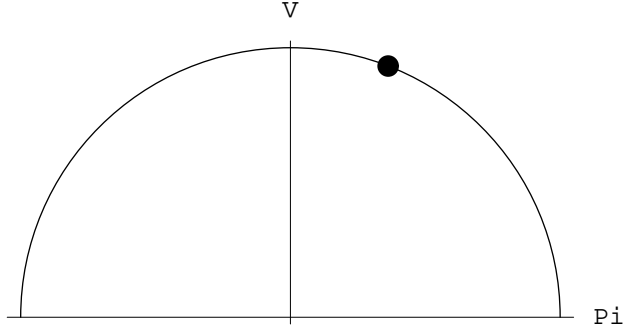}
\end{center}
\caption{The rolling of the tachyon can be described by rolling of a particle down along a semicircle
of radius ${\cal H}$ in the phase space of $V$ and $\Pi.$ }
\end{figure}

Equation (\ref{EM-relation}) can be solved formally as
\begin{equation}
t = \int^T_{T_0}{dT \over \sqrt {1 - {V(T)^2 \over {\cal H}^2}} }, \label{formal-solution}
\end{equation}
where we assumed that the rolling of tachyon starts at $t=0$ with initial value $T_0.$ Even though it is not
easy to perform the integration explicitly, it can easily be done for the following potential
\begin{equation}
V(T) = e^{-{1\over 2} \alpha T}\label{potential-form}
\end{equation}
which is known to be an asymptotic form of the tachyon potential.
Here, $\alpha$ is 1 for bosonic strings, and $\sqrt 2$
for superstrings. The relevant useful relation is the following,
\begin{equation}
\int{ dx \over \sqrt{1 - {1\over b}e^{-ax}} } =  {2\over a}\cosh^{-1}\left\{ \sqrt{b} \exp( {ax\over 2}
)\right\} + {\rm const.}
\end{equation}
The explicit solution is given by
\begin{equation}
T=T_0 + {2\over \alpha}\log\cosh\left( {\alpha\over 2}t\right),\label{the-solution}
\end{equation}
where $T_0,$ when $\Pi(0) \simeq 0,$ is related to ${\cal H}$ by
\begin{equation}
{\cal H} = e^{-{1\over 2}\alpha T_0}.
\end{equation}
\begin{figure}
\begin{center}
\psfig{figure=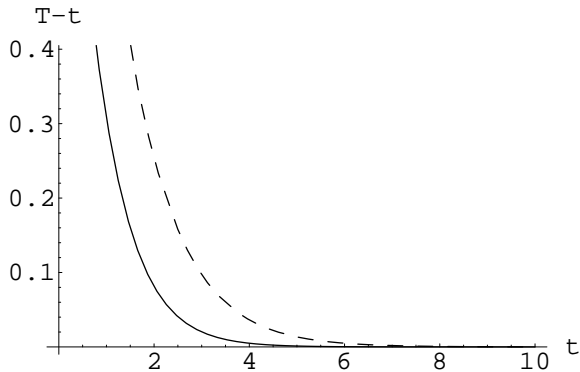}
\end{center}
\caption{The graph which shows that $T$ approaches to $t$ rapidly as $t\to\infty.$ The solid line
corresponds to superstring case, and the dashed line to bosonic string case. In this plot we assumed
${\cal H}={1\over 2}$ for both cases.}
\end{figure}

Using the asymptotic form $\log\cosh x \sim x$ for large $x,$ one finds that $T$ approaches rapidly to $t$
as $t\to\infty.$ In fact it is consistent with the result obtained from Lagrangian
formalism\cite{Lagrangian-formalism}. The advantage of our formalism is that it is more convenient when
applied to a complex case such as a pair of brane-antibrane which will be considered in the next
section.
%
\section{Rolling tachyon solution for a brane-antibrane pair}
%
Instead of homogeneous configuration of tachyon field in space, we consider more realistic one by
assuming the following special form of solution
\begin{equation}
T=a(|x| - \xi) + {\cal T}(t), \;\;\;{\rm for}\;  |x| > \xi. \label{ansatz}
\end{equation}
Here, $x$ is one of spatial $x^i$ coordinate, and $a,$ $\xi$ are finite positive numbers.  This type of
configuration corresponds to the tachyon vacuum for large $|x|.$

It is known from string theoretic calculations\cite{potential-form} that the potential around the
tachyon vacuum looks like (\ref{potential-form}). It is interesting that this type of ansatz, on the
other hand, is valid only if the potential is of type (\ref{potential-form}).  We will prove this fact
later. This means that our ansatz is consistent with the general argument of the potential. For
$|x|<\xi,$ the ansatz may not be valid, and may need a cross term of functions of $x$ and of $t.$ This
term may be related to the possible modification of the potential (\ref{potential-form}) for small $T.$
Physical importance of the ansatz (\ref{ansatz}), according to the interpretation of \cite{Hashimoto}, is
that it corresponds to the tachyon configuration of a brane-antibrane pair.

To solve the equation of motion for the given form (\ref{ansatz}), we start from the energy-momentum
conservation relation (\ref{eq-H}),
\begin{equation}
\partial_0 {\cal H} = \pm a {\partial \Pi \over \partial x},\label{eq-H-pm}
\end{equation}
where $\pm$ corresponds to the sign of $x.$  This can easily be solved when one makes use of
\begin{equation}
{\partial\over \partial x} \left( {\Pi \over V} \right) = 0,
\end{equation}
which comes from (\ref{definition-of-Pi}). Then (\ref{eq-H-pm}) reduces to
\begin{equation}
\partial_0 {\cal H} = \pm a {\Pi \over V} \partial_x V.\label{eq-H-log}
\end{equation}
Using the fact that
\begin{equation}
\partial_x V = \pm a {\partial_0 V \over \partial_0 T}
\end{equation}
one may rewrite (\ref{eq-H-log}) as
\begin{equation}
\partial_0 \log {\cal H} = a^2 {\Pi \over {\cal H} \partial_0 T} \; \partial_0 \log V. \label{d0log}
\end{equation}
This can further be simplified by using
\begin{equation}
{\cal H} \partial_0 T = \Pi ( 1 + (\nabla T)^2)
\end{equation}
which can be proven from either (\ref{H}-\ref{eq_T}) or $Ev =p.$
It means that (\ref{d0log}) reduces to
\begin{equation}
\partial_0 \log\left( {\cal H} V^{ -1 + {1 \over b^2}} \right) = 0, \label{eq-H-gen}
\end{equation}
where $b=\sqrt{1+a^2}.$

From the relation $E =m/ \sqrt{ 1- v^2}$ one may guess that ${\cal H}$ is proportional to $V.$
In fact, by using (\ref{definition-of-Pi}) and (\ref{H}), or equivalently by making use of (\ref{vT}) and (\ref{EH})
in $E =m/ \sqrt{ 1- v^2},$ ${\cal H}$ can be written in terms of $V$
in the following way
\begin{equation}
{\cal H} = {b^2 \over \sqrt{b^2 - \dot{\cal T}^2}}V.
\end{equation}
Then (\ref{eq-H-gen}) reduces to
\begin{equation}
{ \ddot{\cal T}(t) \over b^2 - \dot{\cal T}(t)^2} = -{1 \over b^2}{V'(T)\over V(T)} .\label{ddot-T}
\end{equation}
The left hand side of this equation dependents only on $t,$ while the right hand side of it depends on
$T=a(|x|-\xi) + {\cal T}(t).$ This means that the only possibility is
\begin{equation}
{V'(T)\over V(T)} = -{\alpha \over 2} = {\rm const.} \label{eq-V}
\end{equation}
This shows that the ansatz (\ref{ansatz}) is consistent with the potential of type
(\ref{potential-form}).

By introducing a new variable $z = \dot{\cal T},$ the resulting equation
\begin{equation}
{ \ddot{\cal T}(t) \over b^2 - \dot{\cal T}(t)^2} = {\alpha \over 2 b^2}
\end{equation}
can be written as
\begin{equation}
{1 \over b^2 - z^2} {dz \over dt} = {\alpha \over 2 b^2}. \label{eq-z}
\end{equation}
To solve this we use
\begin{eqnarray}
\int {dx \over 1 - x^2} &=& \tanh^{-1} x + {\rm const},\\
\int  \tanh x \; dx &=& \log\cosh x + {\rm const}.
\end{eqnarray}
The final solution is the following
\begin{equation}
{\cal T} = {\cal T}_0+ {2 b^2 \over \alpha} \log\cosh\left\{{\alpha\over 2b}(t-t_0)\right\},
\end{equation}
where $t_0$ and ${\cal T}_0$ are integration constants.  Note that as $t\to\infty,$ $T \to bt$ which is
quite different from the usual one plotted in Figure 2.

By comparing the result to the homogeneous case, it becomes clear that the decay mode
of the rolling tachyon of a brane-antibrane pair is similar to that of an unstable brane given by (\ref{the-solution}).
The only difference is the characteristic decay time which is enhanced by the factor $b.$
%
\section*{Acknowledgement}
%
This work is supported by the Basic Science Promotion Program of the Chungbuk National University,
BSRI-01-S01.

\end{document}